\input{aipcheck}

\documentclass[
    ,final            
  ]
  {aipproc}

\layoutstyle{6x9}

\begin{document}

\title{Introduction to Nuclear Astrophysics\footnote{Prepared for the Proceedings of the 5th European Summer School on Experimental Nuclear Astrophysics, Santa Tecla, Italy, September 21-26, 2009.}}


\author{Christian Iliadis}{
  address={Department of Physics \& Astronomy, University of North Carolina, Chapel Hill, North Carolina 27599, USA}
}



\begin{abstract}
In the first lecture of this volume, we will present the basic fundamental ideas regarding nuclear processes occurring in stars. We start from stellar observations, will then elaborate on some important quantum-mechanical phenomena governing nuclear reactions, continue with how nuclear reactions proceed in a hot stellar plasma and, finally, we will provide an overview of stellar burning stages. At the end, the current knowledge regarding the origin of the elements is briefly summarized. This lecture is directed towards the student of nuclear astrophysics. Our intention is to present seemingly unrelated phenomena of nuclear physics and astrophysics in a coherent framework. 
\end{abstract}

\maketitle


\section{Introduction}\label{intro}
The field of nuclear astrophysics started with basic questions regarding our Sun.
It is obvious that life on Earth depends on nuclear processes deep inside the Sun, but how exactly the nuclear transmutations occur was not understood for some time. The breakthroughs came at the end of the 1930's: Bethe and Critchfield \cite{Bet38} uncovered a sequential reaction sequence fusing hydrogen (H) to helium (He), now referred to as the ``pp1 chain'', while Bethe \cite{Bet39} and von Weizs\"acker \cite{Wei38} proposed a cyclic reaction sequence, now called the ``CNO1 cycle'', that has the same end result of synthesizing He from H. For this early work, the Nobel prize was awarded to Hans Bethe in 1967. It is interesting to point out that Bethe originally thought that the Sun derives most of its energy via the CNO1 cycle. Part of the problem was that some of the key nuclear reaction cross sections were poorly known. When more reliable cross sections could be estimated in the 1950's, it became apparent that it is in fact the pp1 chain that governs the energy production in the Sun. The important lesson is that accurate nuclear physics information is crucial for our understanding of stars. 

\begin{figure}\label{abunddis}
\includegraphics[height=.3\textheight]{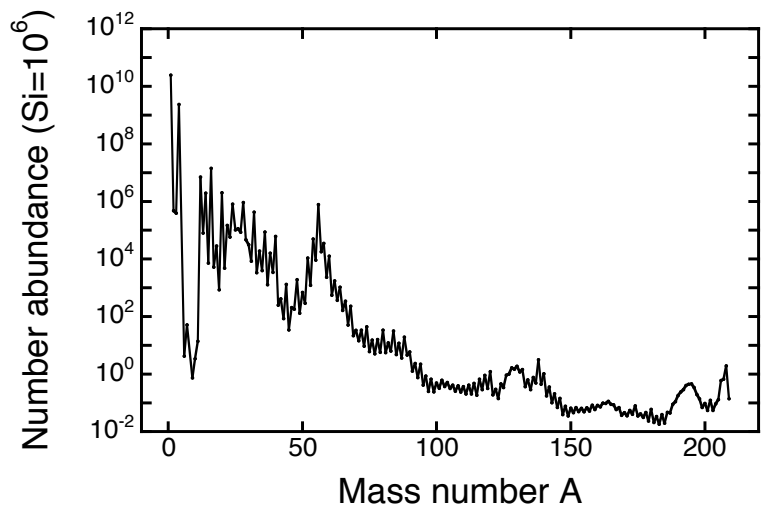}
\caption{Abundances of the nuclides, normalized to the number of Si atoms, at the birth of the solar system. Data from Ref. \cite{Lod03}.}
\end{figure}

Some obvious questions followed immediately: how do other stars produce energy? How do they evolve and why do some of them explode? And perhaps the key question: where were the elements found on Earth produced? They were certainly not produced inside the Sun and, therefore, other processes are required to explain their origin. In this regard the solar system abundance distribution of the nuclides became of paramount importance. It is displayed in Fig. \ref{abunddis} and reveals a rather complicated structure. The different processes giving rise to the observed features were explained by Burbidge, Burbidge, Folwer and Hoyle \cite{Bur57} and by Cameron \cite{Cam57}. These papers laid the foundation of the modern theory of nuclear astrophysics. For this work, the Nobel prize was awarded to Willy Fowler in 1983. 

Briefly, H and He are the most abundant elements and are made in the Big Bang (see contribution of T. Kajino in this volume). The abundance curve then drops sharply by 8 orders of magnitude. The species Li, Be and B are so quickly destroyed inside stars that their production must take place elsewhere. In fact, they are believed to be produced by cosmic-ray spallation (see contribution of J. Kiener). The abundance curve increases sharply at C and O. These are the most abundant elements after H and He and, incidentally, are the species life on Earth is based on. For increasing mass number the abundance curve decreases, but then produces a maximum near Fe, Co and Ni, referred to as the \emph{iron peak}. Interestingly, these nuclides exhibit the largest binding energies per nucleon. So far, most of the species have been produced by nuclear reactions involving charged particles. To explain the origin of the nuclides located beyond the iron peak, however, fundamentally different processes are required. Those species are mainly produced via the capture of neutrons by the \emph{s-process} and the \emph{r-process} (see contributions of M. El Eid and K. L. Kratz, respectively).   

In this lecture, we will focus on charged particle processes important for stellar nucleosynthesis and energy production in stars. It will become apparent how quantum-mechanical processes govern the evolution of large-scale objects in the Universe. This inter-connection is fascinating and remarkable, especially in view of the complex interplay of nuclear physics, hydrodynamics, atomic physics and plasma physics in stars. We may state without exaggerating that after several decades of research, stellar evolution and nucleosynthesis are among the most successful theories humans possess. The account given here is based on a recently published book \cite{Ili07}, to which the student is referred for more detailed information. 

The single most important stellar property that determines the evolutionary fate of a star is its mass. The larger the mass, the larger the temperature and pressure in the core. Thus nuclear energy must be generated at a faster pace in order to stabilize the star, implying a larger luminosity and a faster fuel consumption. Consequently, the larger the mass the shorter the stellar lifetime. Before continuing, it is worthwhile to discuss briefly a few astrophysical phenomena, since they will be mentioned in the following. 

We start with globular clusters. These are located in a spherical space surrounding the Galactic center, called the halo. A typical cluster consists of 10$^4$-10$^6$ stars and is metal poor compared to the Sun, implying that it was formed during the early stages of Galactic evolution. The stars in a single cluster were formed around the same time from material of very similar composition. When plotting the luminosity versus surface temperature for many stars in a given globular cluster (Hertzsprung-Russell diagram), it is apparent that the stars occupy distinct regions in the diagram. This observation must then be explained by differences in  their stellar mass. It is interesting that the age of a cluster can be determined by comparing the location of the turn-off point (i.e., the region in the HR diagram corresponding to those stars that have exhausted the H fuel in their core) with predictions from stellar evolution models, provided that accurate nuclear reaction cross sections are available. Such investigations yield ages for the oldest globular clusters of 12-13 Gy. This value represents an important lower limit for the age of the Universe, demonstrating nicely how nuclear reaction cross sections connect to questions in cosmology. 

It is illuminating to describe briefly the evolution of low-mass stars (M/M$_\odot=0.4-2$), including the Sun. For example, at present the Sun converts H to He in the core via the pp1 chain (see below). In about 5 Gy, the H fuel in the core will be exhausted and the Sun will become a red giant star, fusing H to He via the CNO1 cycle in a shell surrounding an inert He core. The temperature in the core increases until He starts to fuse to C and O (\emph{helium burning}), while H continues to burn in a shell surrounding the core. In this phase, the low-mass stars are referred to as horizontal branch stars. At some point, the He in the core is exhausted and the stars will burn He in a shell surrounding an inert C and O core, in addition to burning H to He in a shell surrounding the He burning region. This phase, referred to as the asymptotic giant branch (AGB), gives rise to thermal instabilities, where the H and He burning shells alternate as the main contributor to the luminosity. As a result, the star will experience a significant mass loss via a strong stellar wind. When the stellar surface becomes hot enough, the intense ultraviolet radiation ionizes the expanding ejecta, which begin to fluoresce brightly as a planetary nebula. Eventually, the H burning shell extinguishes and the low-mass star will end its existence as a white dwarf, consisting mainly of C and O. It is supported by electron degeneracy pressure and cools slowly by radiating away its thermal energy. As will be seen later, AGB stars are believed to be the main sources of C and N in the Universe (and of the \emph{main component} of the s-process).

Massive stars (M$>$11M$_\odot$) evolve very differently from low-mass stars. We will briefly describe the fate of a 25M$_\odot$ star of solar composition. After undergoing H and He burning, the core experiences further burning episodes. These are referred to as C-, Ne-, O- and Si-burning (also called \emph{advanced burning stages}), and will be explained in more detail below. For example, after the end of He burning the core consists of C and O. The core contracts gravitationally, while raising the temperature and pressure, in order to stabilize the star. At some point the conditions are such that C begins to fuse, providing a source of nuclear energy and halting temporarily the contraction of the star. This cycle repeats while the core experiences more such burning stages. Furthermore, each time a given burning phase terminates in the core,
it continues to burn in a shell surrounding the core. The duration of each subsequent nuclear burning phase decreases significantly. For example, while H burning may last many million years, Si burning may last only one day. The reasons are twofold. First, H burning releases far more energy per unit mass (6$\times$10$^{24}$ MeV/g) compared to He and C burning. Hence, the H fuel is consumed much slower. Second, the manner by which the star radiates energy (or ``cools'') changes dramatically. For H and He burning, the nuclear energy generated in the core eventually reaches the surface and is radiated as photons. Beyond He burning, starting with C burning, most of the stars energy is radiated via neutrinos. Since this mechanism of cooling is much more efficient, the fuel consumption increases rapidly.

After the last advanced burning stage, when Si is exhausted,  the core consists mainly of iron peak nuclides (mainly $^{56}$Fe). These nuclides exhibit the largest binding energy per nucleon and thus no more energy can be generated via fusion reactions. In other words, no other source of nuclear energy is available to the star. Meanwhile the mass of the core grows since the overlying burning shells produce more nuclear ashes. When the core grows to a mass near the Chandrasekhar limit (M$\approx$1.4M$_\odot$) electron degeneracy pressure is unable to counteract gravity and the core starts to collapse. Two important effects accelerate the core collapse. First, electrons that could otherwise contribute to the pressure are increasingly removed by electron capture on iron peak nuclei. Second, as will become clear later, with increasing temperature and density the composition of the core shfts to lighter nuclei (which are less stable), thus removing energy and decreasing the pressure. At this stage the core collapses essentially in free fall. When a density of $\approx$10$^{14}$ g/cm$^3$ is reached, the nucleons will start to feel the short-range nuclear force, which is repulsive at very short distances. As a result, the core rebounds and produces an outgoing shock wave. How the shock is precisely generated and how it propagates is not well understood. These topics are subject of current nuclear astrophysics research. For our purposes it is sufficient to state that the outgoing shock heats and compresses the overlying layers of the star, consisting of successive shells of Si, O, Ne and C, before moving outward. Thus the star experiences more episodes of nucleosynthesis, referred to as \emph{explosive Si-, O-, Ne- and C-burning}. It is likely that the precursor star that gave rise to the Crab nebula (M 1) supernova remnant underwent all of these hydrostatic and explosive burning stages referred to above in connection with massive stars.

Our understanding of how single massive stars explode as (core-collapse) type II supernovae has been significantly improved after observation of supernova 1987A (the designation means it was the first supernova discovered in 1987) in the Large Magellanic Cloud. It was the brightest exploding star seen in 400 years. Neutrinos from this event were observed on Earth \cite{Hir87}, providing direct proof of stellar nucleosynthesis. The light curves (luminosity versus time) of many supernovae, including SN 1987A, are powered over an extended time period by the radioactive decay of $^{56}$Co to stable $^{56}$Fe. In fact, $\gamma$-rays from $^{56}$Co decay have even been detected directly for SN 1987A \cite{Mat88}. It will be discussed later how the stellar explosion gives rise to the synthesis of large amounts of radioactive $^{56}$Ni, which is the precursor of $^{56}$Co.

Finally, we will briefly describe a few close binary stellar systems that will be of interest here. Type Ia supernovae are among the most energetic explosions in the universe that sometime even outshine their host galaxies. Their light curves are also powered by the radioactive decay of $^{56}$Co. A detailed understanding of these events is still missing. We will focus here on the most popular model, involving a C-O white dwarf in a close binary star system that accretes matter via Roche lobe overflow from a companion main-sequence or red giant star. The rate of mass accretion must be relatively large ($\approx10^{-7}$M$_\odot$/y) in order to avoid any mass loss through a nova-like event (see below). When the white dwarf grows to a mass near the Chandrasekhar limit, C ignites under degenerate conditions and a thermonuclear runaway occurs. The energy release from nuclear burning is so large that it disrupts the white dwarf, without leaving behind any remnant. For example, SN 1572 (``Tycho's supernova'') is believed to be of type Ia. Since the intrinsic brightness of these events is known within some range, it becomes possible to estimate their distance by measuring the apparent luminosity. Type Ia supernovae can be observed across billions of light years and thus are used as ``standard candles'' for establishing cosmological distances. By recording both their apparent luminosity and their redshifts, observations of very distant type Ia supernovae provide a measure for the expansion history of the Universe. The profound cosmological implications, including evidence for the elusive \emph{dark energy} \cite{Rie98}, provide strong motivation for improving models of type Ia supernovae.

Classical novae are also believed to occur in binary star systems, consisting of a low-mass main-sequence star and a white dwarf. Contrary to type Ia supernovae, however, the accretion rates are much smaller ($\approx10^{-10}$-$10^{-9}$M$_\odot$/y). In this case, matter spirals inward and accumulates on the white dwarf surface, where it is heated and compressed by the strong surface gravity. Hydrogen starts to fuse to He via the pp chains during the accretion phase and the temperature increases gradually. Eventually a thermonuclear runaway occurs, where a significant fraction of the nuclear energy is produced by the \emph{hot CNO cycles}. The luminosity increases by about 4 orders of magnitude. All classical novae are expected to recur every $10^4$-$10^5$ years. The observation of an overabundance of Ne in some classical novae showed that these outbursts do not involve a C-O white dwarf, but a more massive white dwarf of O-Ne composition. The latter objects result from the evolution of intermediate mass stars (M/M$_\odot=9-11$). Other observed overabundances, for example of N, Si and S, are the result of nuclear processing during the explosive burning of H.

Finally, we will discuss type I X-ray bursts. The favorite model involves a low-mass star and a neutron star as the compact object in a binary stellar system. Neutron stars have typical masses near 1.4M$_\odot$, radii of 10-15 km and densities on the order of 10$^{14}$ g/cm$^3$. H- and He-rich matter from the low-mass companion is first accreted in a disk and then falls onto the surface of the neutron star. Temperatures and densities are high enough to fuse H to He via the hot CNO cycles. The accreted He is not fusing yet but sinks deeper into the neutron star atmosphere. Eventually, He starts to fuse under degenerate conditions, triggering a thermonuclear runaway. The burning of a H-He mixture synthesizes elements up to - and perhaps beyond - the iron peak via the \emph{$\alpha$p-process} and the \emph{rp-process}. During a type I X-ray burst the X-ray luminosity increases typically by an order of magnitude. It is unlikely that any matter synthesized during a type I X-ray burst can escape the large gravitational potential of the neutron star. However, these events are important for probing the properties of neutron stars, such as mass, radius and composition.

\section{Nuclear reactions}\label{nucreac}
The cross section of a nuclear reaction is defined as the number of interactions per time, divided by the number of incident particles per area and time, and divided by the number of target nuclei within the beam. The unit is \emph{barn}, where 1 barn$\equiv$10$^{-28}$ m$^2$. For example, the estimated cross section for the reaction $p+p \rightarrow d + e^+ + \nu$, which represents the first step in the pp chains (see below), amounts to $\sigma = 8\times10^{-48}$ cm$^2$ at a laboratory bombarding energy of 1 MeV. Suppose a measurement of this reaction would be performed using an intense 1 mA beam of protons, incident on a dense hydrogen target (10$^{20}$ protons per cm$^2$), then one obtains only 1 interaction in 6000 years! Clearly, such a measurement is beyond present experimental capabilities and hence this cross section needs to be estimated theoretically.

Cross section curves (cross section versus bombarding energy) come in many varieties. In the simplest case, the cross section of a charged-particle reaction drops dramatically with decreasing energy, but otherwise 
exhibits no structure. A good example is the cross section for $^{16}$O(p,$\gamma$)$^{17}$F below a center of mass energy of 2 MeV. Sometimes the cross section exhibits a well-defined maximum. An example for such a behavior is the $^{13}$C(p,$\gamma$)$^{14}$N reaction, which shows a maximum near 500 keV in the center of mass. It is frequently stated that the Coulomb barrier is responsible for the sharp drop in cross section with decreasing energy, while the cross section maxima are identified as \emph{resonances}. But how does the Coulomb barrier exactly explain the observation? And what is the origin of a resonance?

\begin{figure}\label{pot}
\includegraphics[height=.3\textheight]{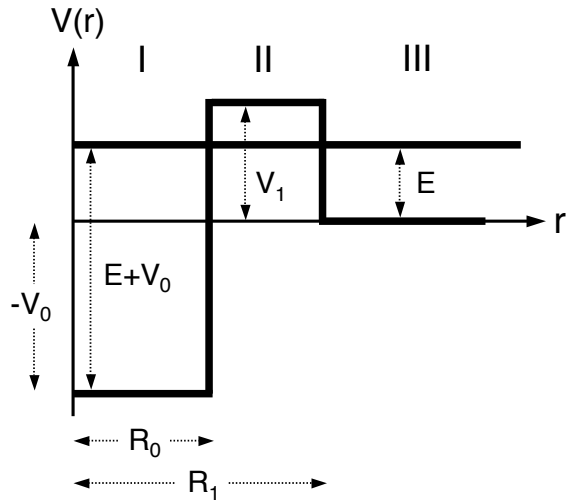}
\caption{Three-dimensional square-well-plus-square-barrier potential, representing the simplest potential model for a charged-particle nuclear reaction. The potential properties are defined in the text. Note that for a real potential the transmission probability is only defined for the one-dimensional case. Thus for the calculation of the transmission coefficient we will assume that the figure represents a one-dimensional potential that extends to $-\infty$.}
\end{figure}

These questions can best be answered by considering a simple potential model for a nuclear reaction. We need essentially two pieces: an attractive (negative) square well potential of depth $V_0$ inside the nucleus ($r<R_0$), and a repulsive (positive) square barrier potential of height $V_1$ for distances of $R_0\leq r < R_1$. The total energy of the incident particle, $E$, is less than the barrier height. An example for such a potential is shown in Fig. \ref{pot}. The solution to the Schr\"odinger equation for each of the three regions I, II, III are well know and can be found in any introductory quantum mechanics textbook. In regions I and III, the solutions are in the form of complex exponentials, which represent a sine function. In region II, however, the solutions are given in terms of real exponentials. In the next step, the \emph{continuity condition} is applied, that is, the wave function solutions and their derivatives must be continuous at the two boundaries $R_0$ and $R_1$. We obtain four equations and can solve for the intensity of the transmitted wave in region I, normalized to the intensity of the incident wave in region III. For relatively low energies one finds for this ratio, also called \emph{transmission coefficient}, after some algebra 
\begin{equation}
\hat T \approx e^{-(2/\hbar) \sqrt{2m(V_1 - E)} (R_1 - R_0)}\label{tunn}
\end{equation}
where $m$ denotes the reduced mass. This result, referred to as the \emph{Tunnel effect}, is remarkable and represents one of the most important quantum mechanical phenomena for charged particle reactions: although the incoming particle is classically not allowed to reach region I, there is in fact a finite probability for tunneling through the barrier. Without this circumstance the world would be a very different place and life on Earth would certainly not exist. It is clear from Eq. (\ref{tunn}) that the transmission coefficient depends very sensitively on the properties of the barrier. Its energy dependence, especially the sharp drop with decreasing energy, broadly resembles that of the $^{16}$O(p,$\gamma$)$^{17}$F cross section.

We have not considered yet the full radial wave function solution of the Schr\"odinger equation for the three-dimensional case. In particular, we are interested in the ratio, $R$, of the wave function intensities in regions I and III. It is now of advantage to express the wave function solutions in these regions as sine functions instead of complex exponentials. Otherwise we proceed as before: we apply the continuity condition and solve the system of four equations for the ratio R. After some tedious algebra we find a rather lengthy analytical expression. Interestingly, when plotting this function versus energy $E$ for certain values of the potential depth $V_0$, a well-defined maximum is produced, while for other values of $V_0$ the resulting function closely reproduces the structureless energy-dependence of the transmission coefficient in Eq. (\ref{tunn}). Note that by changing the potential depth $V_0$ we are changing the wavelength in the nuclear interior (region I), and for discrete values of $V_0$ the wave function amplitude in the nuclear interior becomes relatively large. This describes, in the simplest case, the origin of a well-defined cross section maximum: \emph{a resonance results from favorable wave function matching conditions at the nuclear boundary}. 

Obviously, in a more realistic situation we need to replace the simple square barrier by the Coulomb potential (if we disregard the centripetal barrier for the moment). The Coulomb potential has a more complicated shape, but we may approximate it by dividing this potential into many thin square barriers. The transmission coefficient for the Coulomb potential is then given by the product of the transmission coefficients for all of the square barriers. If we let the number of square barriers become very large ($n\rightarrow \infty$), the transmission coefficient for the Coulomb potential can be found analytically. For very low energies we find
\begin{equation}
\hat T \approx \exp\left(- \frac{2\pi}{\hbar}\sqrt{\frac{m}{2E}} Z_0 Z_1 e^2\right)  \equiv e^{- 2\pi \eta} \label{gamowfacexpr}
\end{equation}
where $Z_0$ and $Z_1$ are the charges of the interacting nuclei and $e$ is the elementary charge. This function reveals a $1/\sqrt{E}$ dependence in the exponent and is referred to as the \emph{Gamow factor}. It is frequently used in nuclear astrophysics to define a rather useful quantity, called the astrophysical S-factor, via the relation $S(E)\equiv E\sigma(E)\exp(2\pi\eta)$: division by the Gamow factor removes from the cross section $\sigma(E)$ the strong Coulomb barrier transmission probability and produces a function $S(E)$ that is more manageable (for example, in theoretical extrapolations to very low energies). 

In formal reaction theory, a simple equation describing a single isolated resonance can be derived. It is referred to as \emph{Breit-Wigner formula} and is given by
\begin{equation}
\sigma_{\mathrm{BW}}(E) = \frac{\lambda^2\omega}{4\pi} \frac{\Gamma_a \Gamma_b }{(E_r - E)^2 + \Gamma^2/4}\label{breitW}
\end{equation}
where $\lambda$ is the de Broglie wavelength, $\omega$ is a factor containing angular momenta, $E_r$ is the resonance energy, $\Gamma_i$ are the resonance partial widths of entrance and exit channel, and $\Gamma$ is the total resonance width given by the sum of all partial widths. The above equation is the single most important expression describing a resonance and it is frequently used in nuclear astrophysics in many applications: (i) for fitting cross section data to extract resonance parameters; (ii) for deriving the ``narrow-resonance reaction rate'' (see below); (iii) for extrapolating cross sections to energy regions were no measurements exist; and (iv) for calculating the experimental resonance yield when the resonance cannot be resolved experimentally.

A partial width describes the probability (in energy units) per unit time for formation or decay of a resonance. For example, the partial width for forming a resonance via proton absorption, or for decay of a resonance via proton emission, is given by the expression
\begin{equation}
\Gamma_{\lambda c} = 2\frac{\hbar^2}{mR^2} P_c\,C^2 S\,\theta_{pc}^2\label{partW}
\end{equation}
Apart from a constant factor involving the reduced mass $m$ and the nuclear radius $R$, the proton partial width is given by the product of three distinct probabilities: first, the probability that nucleons will arrange themselves in a ``target-plus-single-particle'' configuration (\emph{spectroscopic factor}, $C^2S$); second, the probability that a proton will appear at the nuclear boundary (\emph{dimensionless reduced single-particle width}, $\theta_{pc}^2$); and, finally, the probability that the proton will penetrate the Coulomb and centripetal barriers (\emph{penetration factor}, $P_c$). The third factor, which can be computed precisely from Coulomb wave functions, is strongly energy-dependent. The second factor can also be computed numerically \cite{Ili97}.
The great untility of Eqs. (\ref{breitW})-(\ref{partW}) becomes now apparent: if the spectroscopic factor, $C^2S$, which is a nuclear structure quantity, can be estimated by different means, for example, using transfer reaction studies, then the partial width can be calculated and the cross section be estimated in a straightforward manner, despite the fact that the reaction cross section has not been measured directly. Clearly, in many cases the cross section cannot be measured directly, either because the Coulomb barrier transmission probability is too small or perhaps because the target is short-lived. Consequently, such \emph{indirect methods} of estimating the cross section become a crucial tool in nuclear astrophysics (see contributions of A. Mukhamedzhanov, G. Rogachev, and L. Trache). 

\section{Thermonuclear reactions}\label{thermoreac}
In a stellar plasma, the kinetic energy for a nuclear reaction derives from the thermal motion of the participating  nuclei. Hence, the interaction is referred to as \emph{thermonuclear reaction}. The thermonuclear reaction rate (the number of reactions per unit time and unit volume) for a reaction $0+1\rightarrow 2+3$ is given by $r_{01}=N_0N_1\left< \sigma \mathrm{v} \right>_{01}$, where $N_i$ are the number densities of the interacting nuclei and $\left< \sigma \mathrm{v} \right>_{01}$ is the reaction rate per particle pair, which is equal to the integral over the product of velocity, cross section and velocity probability density. In most cases of practical interest, the latter function is given by the Maxwell-Boltzmann distribution. Thus the reaction rate per particle pair can be written as
\begin{equation}  
\langle \sigma \mathrm{v} \rangle_{01} = \left(\frac{8}{\pi m_{01}}\right)^{1/2} \frac{1}{(kT)^{3/2}} \int_0^\infty E\,\sigma(E)\,e^{-E/kT}\,dE \label{rate}
\end{equation}  
where $m_{01}$ is the reduced mass, $k$ the Boltzmann constant and $T$ the plasma temperature. Clearly, for a given temperature the reaction rate is precisely determined if the nuclear reaction cross section, $\sigma(E)$, is known.  

At this point it is worthwhile to note that a given nuclear reaction occurring in the stellar plasma can rarely be considered as an isolated interaction. Consider, for example, the species $^{25}$Al at an elevated temperature. It may be destroyed in several different ways: via $\beta^+$-decay to $^{25}$Mg, via proton capture to $^{26}$Si, via photodisintegration to $^{24}$Mg, and so on. On the other hand, $^{25}$Al is produced via the $\beta^+$-decay of $^{25}$Si, via proton capture on $^{24}$Mg, via photodisintegration of $^{26}$Si, and so on. The abundance evolution of $^{25}$Al during nucleosynthesis is then given by a differential equation that accounts for all destruction and production mechanisms. Of course, such a differential equation needs to be written for all species participating in the nuclear burning. Thus one ends up with s system of coupled differential equations, called a \emph{nuclear reaction network}. A good introduction of how to solve this system numerically can be found in Arnett's book \cite{Arn96}. 


It is interesting to investigate Eq. (\ref{rate}) in more detail by considering a few extreme examples. We start with the simplest case, i.e., a nearly constant S-factor, $S_0$. This situation is usually referred to as ``non-resonant'', which however leads to considerable misunderstandings since the formalism also applies to slowly varying resonance ``tails'', as will be seen below. Substitution of the S-factor definition (see above) into Eq. (\ref{rate}) shows immediately that the reaction rate depends, apart from the magnitude of $S_0$, on the integral over the product of Gamow and Boltzmann factors, $e^{-2\pi\eta}e^{-E/kT}$. This result is significant because it demonstrates that the star does not burn at high energies where the cross section is large (since the number of particles with such energies is vanishingly small); neither does the star burn at very small energies where the number of particles is at maximum (since the cross section is vanishingly small). Rather, in a plasma most nuclear reactions occur at energies where the function $e^{-2\pi\eta}e^{-E/kT}$ is at maximum. This well-defined energy window is referred to as the \emph{Gamow peak}. 

When the Gamow peak is plotted for a given temperature, but for different target-projectile combinations (implying different projectile and target charges and hence different Coulomb barrier heights), a few important observations can be made. For increasing charges $Z_0$ and $Z_1$: (i) the Gamow peak shifts to higher energies; (ii) the Gamow peak becomes broader; and most importantly, (iii) the area under the Gamow peak decreases dramatically. In other words, for a mixture of different nuclides in a stellar plasma at given temperature, those reactions with the smallest Coulomb barrier produce most of the energy and are consumed most rapidly. This is of paramount importance for the star since it explains the occurrence of well-defined 
stellar burning stages.

Next, we will consider a ``narrow resonance''. Several different definitions for a narrow resonance can be found in the literature, but none of them is without problems. For the sake of simplicity, let us assume that a narrow resonance implies constant partial widths over the total width of the resonance. Substitution of Eq. (\ref{breitW}) into Eq. (\ref{rate}) yields immediately $\left< \sigma \mathrm{v} \right>=[(2\pi)/(mkT)]^{3/2}\hbar^2e^{-E_r/kT}\omega\gamma$. The product $\omega\gamma\equiv \omega\Gamma_a\Gamma_b/\Gamma$ is proportional to the area under the narrow-resonance cross section curve and thus is called \emph{resonance strength} (with units of energy). Note that the resonance energy enters exponentially in the above reaction rate expression. It needs to be determined rather precisely, otherwise the resulting uncertainty of the reaction rate becomes relatively large. 

In many cases the energy-dependence of the partial widths over the total width of the resonance cannot be disregarded. Such ``broad resonance'' reaction rates need to be treated with care. Although approximate expressions exist in the literature, it is safer to substitute Eq. (\ref{breitW}) into Eq. (\ref{rate}) and evaluate the integral numerically. Depending on the location of the broad resonance with respect to the Gamow peak, there are in general two contributions to the total reaction rate. First, the contribution calculated from the ``narrow resonance'' reaction rate, which arises only from the region near the resonance energy (as is apparent from the factor $e^{-E_r/kT}$). Second, from the smoothly varying tail of the resonance. If the broad resonance is located outside the Gamow peak, then in most cases the resonance tail makes a far larger contribution than what is calculated from the narrow resonance expression. Plotting such reaction rates versus temperature frequently reveals a ``kink'' because the narrow resonance and broad resonance reaction rates have different temperature dependences. 

Generally, in order to evaluate the total rate of a single reaction, many different contributions need to be taken into account: narrow and broad resonances, non-resonant processes, subthreshold resonances, cross section continua, interferences between different amplitudes, and so on. Every single reaction represents a special case and the evaluation process is usually tedious. Evaluations of reaction rates have been provided by W. Fowler and collaborators for many years, with their last evaluation (covering the A=1-30 target mass range) published in 1988 \cite{CF88}. A European effort, by the NACRE collaboration, resulted in an updated reaction rate evaluation in 1999 \cite{Ang99}, while another evaluation including for the first time radioactive target nuclei was published in 2001 \cite{Ili01}. As of this writing, we have submitted for publication a major new reaction rate evaluation utilizing Monte Carlo methods for the first time \cite{Ili09}.

\section{Stellar burning stages}\label{stellburn}
\subsection{Hydrostatic hydrogen burning}
\begin{figure}\label{Hburn}
\includegraphics[height=.67\textheight,angle=-90]{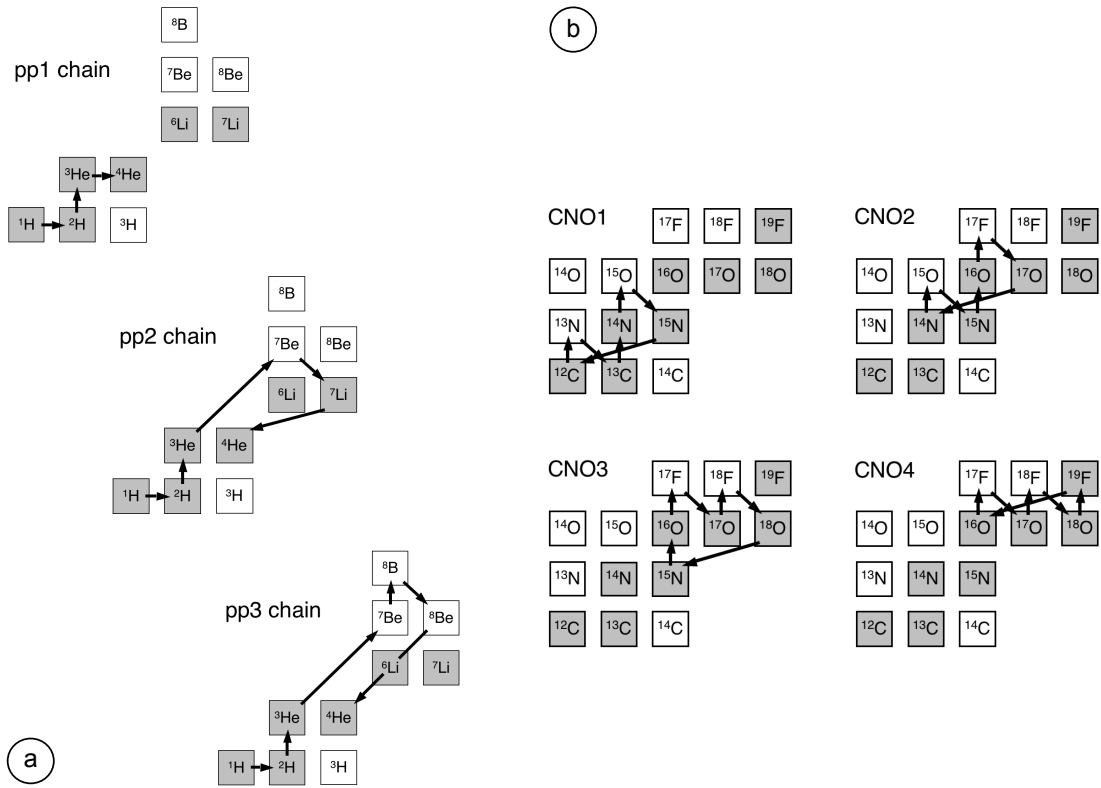}
\caption{The pp chains (part a) and the CNO cycles (part b) shown schematically in the chart of the nuclides. Each arrow represents a specific interaction, connecting an initial with a final nuclide. Stable nuclides are shown as shaded squares. The proton and neutron numbers increase in the vertical and horizontal directions, respectively.}
\end{figure}

Hydrostatic burning of H occurs near $T=15.6$ MK in the center of the Sun, in the range of $T=8-55$ MK in the cores of other stars depending on their mass, and at $T=45-100$ MK in the H burning shell of AGB stars. If only H and He are available as fuel, without the presence of heavier nuclides, then the stellar core generates nuclear energy via the \emph{pp chains}. These are shown schematically in Fig. \ref{Hburn}a. All chains fuse effectively four protons to one $^{4}$He nucleus and thereby generate an energy of 26.7 MeV. Furthermore, at low temperatures all chains involve non-resonant reactions only. Each chain starts with the p(p,e$^+\nu$)d reaction, which has not been measured directly at the relevant energies (see above). The absolute magnitude of this cross section is influenced by the weak interaction. Fortunately, the different factors that determine the S-factor can be estimated theoretically with substantial confidence \cite{Bac69}. The present reaction rate uncertainty amounts to a few percent only \cite{Ang99}, which is significantly smaller compared to rate uncertainties of most directly measured stellar fusion reactions. Several other reactions that are part of the pp chains have been measured directly, most recently the d(p,$\gamma$)$^3$He, $^{3}$He($^{3}$He,2p)$\alpha$ and $^{3}$He($\alpha$,$\gamma$)$^{7}$Be reactions by the LUNA collaboration (see contribution of G. Imbriani in this volume). 

In many situations small amounts of $^{12}$C and $^{16}$O will be present in the stellar plasma and these nuclei will participate in the stellar burning. For example, $^{12}$C captures a proton yielding $^{13}$N, which in turn, $\beta$-decays to $^{13}$C. This nuclide captures another proton, and then another one to yield $^{15}$O, which $\beta$-decays to $^{15}$N. At this point something important occurs: rather than capturing another proton, $^{15}$N prefers to undergo a (p,$\alpha$) reaction, producing again $^{12}$C. The interesting point here is that by completing one cycle, four protons have fused to one $^{4}$He nucleus, while the heavy ``seed'' nucleus has been recovered. Thus, $^{12}$C acts as a catalyst and even small amounts of CNO material can give rise to a large nuclear energy generation. A small leakage of material via $^{15}$N(p,$\gamma$)$^{16}$O initiates more cyclic reaction sequences. They are referred to as \emph{CNO cycles} and are shown in Fig. \ref{Hburn}b. In each case, the (p,$\alpha$) reaction is favored over the (p,$\gamma$) reaction at the branching points $^{15}$N, $^{17}$O, $^{18}$O and $^{19}$F, which is a necessary condition for a reaction cycle to occur. Obviously, the first cycle, called \emph{CNO1 cycle}, is the most important one. It is governed by $^{14}$N(p,$\gamma$)$^{15}$O, since it is by far the slowest interaction among the reactions and $\beta$-decays. This reaction has been measured both at the LUNA facility (deep underground in Gran Sasso) \cite{LUN}, and at the LENA facility (at sea-level in our laboratory) \cite{LEN}. It was found that the new reaction rate deviates from the previous one \cite{Ang99} by about a factor of 2. As a consequence, the ages of globular clusters, obtained by fitting the turn-off point in the Hertzsprung-Russell diagram to stellar models of low-mass stars (see above), changed by about 1 Gy! This again emphasizes the dramatic impact of precise cross section measurements on stellar models and on cosmological questions. The solar carbon isotopic number abundance ratio is $^{13}$C/$^{12}$C=0.01, while the CNO1 cycle equilibrium ratio amounts to  $^{13}$C/$^{12}$C=0.25. Many stars that burn H via the CNO1 cycle have observed ratios between these two values, while a few stars even come close to the equilibrium value. This implies that a significant fraction of these stars' hydrogen envelope has been cycled through regions that experience equilibrium operation of the CNO1 cycle. From the latest reaction cross sections, assuming a solar composition, one finds that near $T=20$ MK the CNO1 cycle takes over from the pp1 chain as the main energy-generating process. Thus it is found that about 90\% of the Sun's energy generation originates from the pp1 chain. Note that the CNO cycles occurring in AGB stars are predicted to be the main source of $^{13}$C and $^{14}$N in the Universe.

\subsection{Explosive hydrogen burning}
\begin{figure}\label{Hexp}
\includegraphics[height=.65\textheight,angle=-90]{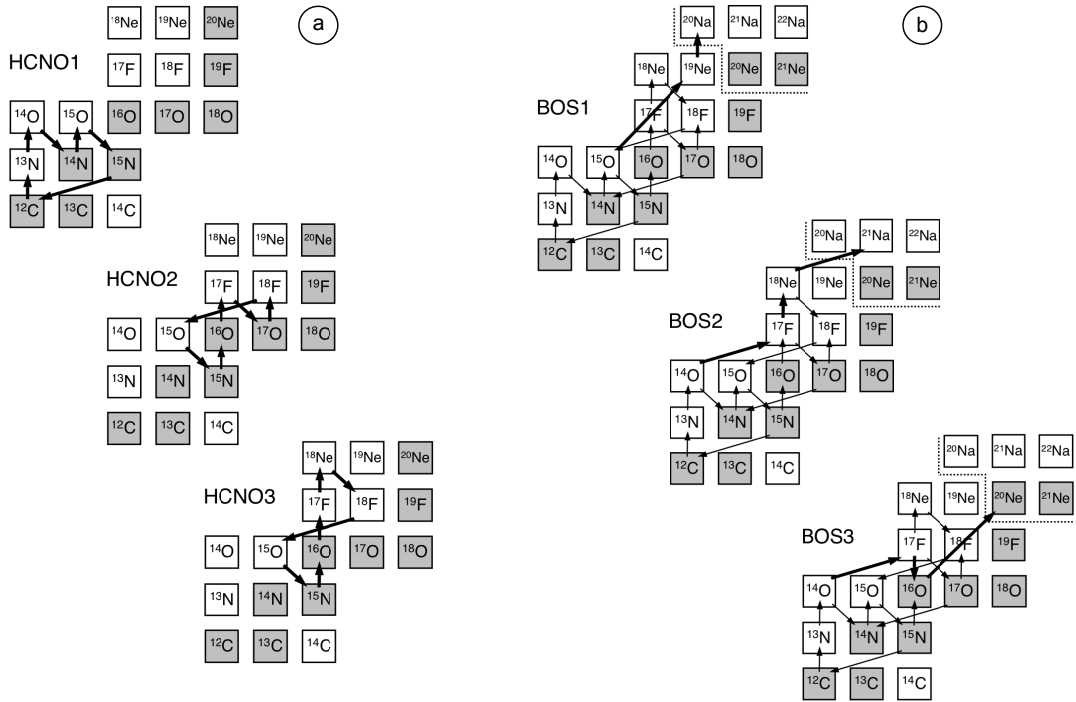}
\caption{The hot CNO cycles (part a) and the breakout sequences (thick arrows in part b).}
\end{figure}

Explosive burning of H in a classical nova outbursts attains much higher peak temperatures ($T=0.1-0.4$ GK) compared to the hydrostatic scenarios discussed so far. At such temperatures the character of the burning changes significantly. The major reactions that occur in the $A<20$ mass range are referred to as the \emph{hot CNO cycles} and are displayed in Fig. \ref{Hexp}a. Consider first the HCNO1 cycle: at these higher temperatures, $^{13}$N prefers to capture a proton, yielding $^{14}$O, instead of $\beta$-decaying to $^{13}$C, as was the case for the (cold) CNO1 cycle (see above). Other than this change the HCNO1 cycle in Fig. \ref{Hexp}a looks just like the CNO1 cycle in Fig. \ref{Hburn}b. However, the inner workings of these cycles are entirely different. The temperatures are now so high that all reactions, including $^{14}$N(p,$\gamma$)$^{15}$O, occur on much shorter time scales than the $\beta$-decays. Thus it is the $\beta$-decays of $^{14}$O and $^{15}$O that control the energy generation rate and cause $^{14}$O and $^{15}$O to be the most abundant nuclides. For this reason the HCNO1 cycle is sometimes referred to as ``$\beta$-limited CNO cycle''. The time to complete the cycle is given by the sum of the mean lifetimes of $^{14}$O and $^{15}$O, which amounts to about 300 s. Since a classical nova explosion only lasts for a few 100 s, it is obvious that the HCNO1 cycle in these events must operate far from equilibrium. 

The other cycles, HCNO2 and HCNO3, are very important for the nucleosynthesis since they explain the fate of the initially abundant $^{16}$O. In particular, reaction network calculations showed that the $^{17}$O+p and $^{18}$F+p reactions are of crucial importance since they influence a number of classical nova observables, for example, the amount of long-lived radioactivity produced (with the hope of measuring decay $\gamma$-rays using satellite-borne instruments), oxygen isotopic ratios (with the hope of measuring these in presolar grains that originate from classical novae; see contribution of P. Hoppe), and the Galactic synthesis of the species $^{17}$O. The $^{17}$O(p,$\gamma$)$^{18}$F and $^{17}$O(p,$\alpha$)$^{14}$N reactions have been measured recently by our group \cite{Fox05} and by the Orsay group \cite{Cha07}. Curiously, it is found that a non-resonant reaction mechanism (direct capture) determines the total reaction rates for $^{17}$O(p,$\gamma$)$^{18}$F at $T=0.1-0.4$ GK, despite the fact that a narrow resonance is located inside the Gamow peak. Clearly, this represents a rather rare case. A measurement of the direct capture process has just been completed by our group \cite{New09}. A direct study of the important $^{18}$F(p,$\alpha$)$^{15}$O reaction requires a radioactive $^{18}$F beam. Despite several measurements, there is still crucial information missing for calculating reliable rates of this reaction.

The nucleosynthesis in classical novae not only takes place in the $A<20$ mass range, but also involves many reactions in the $A\geq20$ range. The details are rather complex. A quantitative investigation of the importance of specific reactions to the overall nucleosynthesis can be found in Ref. \cite{Ili02}, where about 7000 numerical reaction network computations have been performed in a reaction rate sensitivity study.

\subsection{Hydrostatic helium burning}
Hydrostatic burning of helium, for example, in massive stars takes place in the temperature range $T=0.1-0.4$ GK. Helium burning starts with the fusion of two $\alpha$-particles. However, the composite nucleus $^{8}$Be lives for only $\approx10^{-16}$ s and decays back into two $\alpha$-particles. Nevertheless, after a given time a tiny equilibrium abundance of $^{8}$Be builds up, sufficient to allow for capture of a third $\alpha$-particle to form stable $^{12}$C. This process is referred to as \emph{triple-$\alpha$ reaction}. It was pointed out by Hoyle \cite{Hoy54} that this process would be too slow to account for the fusion of $^{12}$C, unless a resonance exists right above the $^8Be + \alpha$ threshold, which furthermore must be formed without inhibition by the centripetal barrier (i.e., it has to be a s-wave resonance). A few years later this level in $^{12}$C, referred to since as the \emph{Hoyle state}, was experimentally verified. The prediction and the subsequent verification of this state represents a marvelous interplay of astrophysics and nuclear physics. The triple-$\alpha$ reaction is a (sequential) three-body interaction and thus has not been measured in the laboratory. From experimental knowledge of the nuclear masses and partial widths involved, the reaction rate can be estimated fairly accurately. Present uncertainties amount to about $\pm$15\% \cite{Ang99}, a remarkably small value for a process that has not be measured directly. 

Helium burning continues via the $^{12}$C($\alpha$,$\gamma$)$^{16}$O reaction. There is no resonance near and above the $\alpha$-particle threshold in $^{16}$O and thus this process must proceed via broad-resonance tails (including subthreshold resonances) and direct mechanisms. These amplitudes may interfere, causing problems in the extrapolation of the S-factor to the astrophysically important energy range, which at present is not accessible experimentally. The rate of this reaction is of great importance, since it determines the $^{12}$C to $^{16}$O abundance ratio at the end of helium burning. This abundance ratio sensitively influences not only all the subsequent hydrostatic burning stages in massive stars, but also the explosive burning, and the nature of the remnant left behind after the core collapse. At present the reaction rate is uncertain by $\pm$35\%, to be conservative, but a more accurate rate is highly desirable. The status of the available data and extrapolations to relevant energies will be discussed in the contribution by C. Brune. The subsequent $^{16}$O($\alpha$,$\gamma$)$^{20}$Ne reaction is very slow, thus explaining the survival of $^{16}$O during helium burning. The main products at the end of helium burning are $^{12}$C to $^{16}$O.

If $^{14}$N is present in the stellar plasma (from CNO cycle operation during the preceding H burning state), then the reaction sequence $^{14}$N($\alpha$,$\gamma$)$^{18}$F($\beta^+\nu$)$^{18}$O($\alpha$,$\gamma$)$^{22}$Ne can be initiated. The subsequent $^{22}$Ne($\alpha$,n)$^{25}$Mg reaction is the major neutron source towards the end of helium burning, and gives rise to the \emph{weak component of the s-process} (see contribution by M. El Eid). Hydrostatic helium burning is predicted to be the main source of $^{12}$C, $^{16}$O, $^{18}$O and $^{22}$Ne in the Universe. 

\subsection{Explosive hydrogen-helium burning}
So far, we only considered the burning of a pure fuel (either H or He). There are situations, however, where the temperatures are high enough to ignite a H-He mixture explosively. An example are type I X-ray bursts (see introduction) that may achieve peak temperatures near $T=1.5$ GK. Such elevated temperatures, together with the presence of He, change entirely the character of the nuclear burning. The hot CNO cycles do not operate any longer, but rather reaction sequences involving both proton and $\alpha$-particle induced reactions initiate a breakout from the CNO mass region to heaver nuclides. The most likely breakout sequences (BOS) are shown in Fig. \ref{Hexp}b as thick lines. Once these sequences reach nuclides beyond the A=20 demarkation (dashed line), there is no likely process that can return matter to the CNO mass range. Consider, for example, the first breakout sequence, $^{15}$O($\alpha$,$\gamma$)$^{19}$Ne(p,$\gamma$)$^{20}$Na. It is apparent that these reactions involve short-lived target nuclei. The same is true for the other sequences (with the exception of the $^{16}$O($\alpha$,$\gamma$)$^{20}$Ne reaction in BOS3). Hence, their experimental study requires  radioactive ion beams. At present, the reactions $^{15}$O($\alpha$,$\gamma$)$^{19}$Ne, $^{14}$O($\alpha$,p)$^{17}$F, $^{17}$F(p,$\gamma$)$^{18}$Ne and $^{18}$Ne($\alpha$,p)$^{21}$Na are subject of ongoing research (see contributions by L. Trache and T. Aumann). 

The breakout sequences are significant because they initiate a network involving numerous reactions on the proton-rich side of the nuclidic chart that may stretch up to $^{68}$Se and beyond. The details are quite complex. In brief, a series of ($\alpha$,p) reactions quickly converts matter up to the Ar region (the $\alpha$p-process), where sequences of (p,$\gamma$) reactions and $\beta^+$-decays take over (the rp-process). At several locations the nuclear activity extends all the way, and sometimes beyond, the line of particle-unstable nuclides (the so-called \emph{proton dripline}). Frequently, when this happens, a forward ($A\rightarrow B$) and reverse ($B\rightarrow A$) reaction are in \emph{equilibrium}, that is, the rates of the forward and reverse reaction ($A\Leftrightarrow B$) become equal. A sequential link, $B\rightarrow C$ , either reaction or $\beta$-decay, usually provides a leak out of the equilibrium. It is very important to note that, once equilibrium has been achieved between species $A$ and $B$, the rates of the reactions $A\rightarrow B$ and $B\rightarrow A$ are entirely irrelevant for the nuclear transformations. This can be shown in a straightforward way by applying the Saha statistical equation and the reciprocity theorem of nuclear reactions \cite{Ili07}. All that is needed (apart from some less important factors) is the $Q$-value of the $A\rightarrow B$ reaction and the reaction rate for the ``leakage'' $B\rightarrow C$ process. For type I X-ray bursts, the two most important equilibria occur at $^{64}$Ge$\Leftrightarrow^{65}$As$\rightarrow^{66}$Se and $^{68}$Se$\Leftrightarrow^{69}$Br$\rightarrow^{70}$Kr. At present the reaction Q-values of $^{64}$Ge(p,$\gamma$)$^{65}$As and $^{68}$Se(p,$\gamma$)$^{69}$Br are rather uncertain, resulting in significant uncertainty regarding the end point and the ashes of the nucleosynthesis in type I X-ray bursts \cite{Par09}.  

\subsection{Hydrostatic carbon, neon, oxygen and silicon burning}
As we have seen, after the end of He burning the core consists mainly of $^{12}$C and $^{16}$O. Thus the next most likely nuclear fuel to ignite is $^{12}$C. For the first time in the life of the star, a heavy-ion fusion reaction, $^{12}$C+$^{12}$C, is defining a burning stage, referred to as \emph{carbon burning}. Typical core temperatures amount to $T=0.6-1.0$ GK. There are three possible \emph{primary reactions}, $^{12}$C($^{12}$C,p)$^{23}$Na,  $^{12}$C($^{12}$C,$\alpha$)$^{20}$Ne and  $^{12}$C($^{12}$C,n)$^{23}$Mg. The released light particles undergo several \emph{secondary reactions} involving newly formed nuclei, among them $^{25}$Mg(p,$\gamma$)$^{26}$Al that will be mentioned again later. The main ashes of C burning are $^{16}$O, which has not participated much in the nuclear activity, and $^{20}$Ne. The primary reaction, $^{12}$C+$^{12}$C, populates levels in the compound nucleus $^{24}$Mg near 14 MeV excitation energy. This energy region exhibits a very high level density, with many broad and overlapping states. Therefore, we expect the S-factor to be a smooth function of energy. However, experiments have uncovered many sharp maxima in the S-factor curve, even near astrophysically important energies. A satisfactory reaction model to explain this structure is lacking at present. This is a problem since the available data do not cover the entire astrophysically important region and hence we have to rely on rather uncertain extrapolations. The present status of this reaction will be discussed in more detail in the contribution of F. Strieder.

After C burning, when the temperature in the core reaches values of $T=1.2-1.8$ GK, the most likely process to occur is the photodisintegration of $^{20}$Ne via the (primary) $^{20}$Ne($\gamma$,$\alpha$)$^{16}$O reaction. For the first time in the life of the star a photodisintegration defines a burning stage. The released $\alpha$-particles (including proton and neutrons at a slightly later time) initiate a number of secondary reactions and the evolving reaction network is referred to as \emph{neon burning}. Although the primary reaction is endothermic (it consumes energy), together with the secondary reactions there is a net production of nuclear energy for each $^{20}$Ne nucleus destroyed. The main nuclear ash of Ne burning is $^{16}$O.

The core contracts further after Ne burning until temperatures of $T=1.5-2.7$ GK are produced. At this stage 
another (primary) heavy ion reaction, $^{16}$O+$^{16}$O, initiates a burning stage, called \emph{oxygen burning}. The temperatures are so high that many exit channels are open, some of which even involve the emission of three particles. The light particles then initiate a number of secondary reactions. It is interesting to note that unlike the case of $^{12}$C+$^{12}$C, the $^{16}$O+$^{16}$O reaction exhibits, as expected a smooth energy-dependence of the S-factor. Nevertheless, the S-factor data at the lowest measured energies are in poor agreement and, furthermore, the branching ratios for the different exit channels need also be known to better accuracy. Clearly, more laboratory work is required. The main ashes of O burning are $^{28}$Si and, to a somewhat lesser extent, $^{32}$S.

After O burning the core contracts until temperatures of $T=2.8-4.1$ GK are reached. At this point the photodisintegration $^{28}$Si($\gamma$,$\alpha$)$^{24}$Mg initiates another burning stage, called \emph{silicon burning}. As was the case before, the released light particles give rise to a network of secondary reactions, but on a much larger scale than during Ne burning. In essence, during this photodisintegration rearrangement, less tightly bound nuclides are photodisintegrated and the released protons, neutrons and $\alpha$-particles are captured to synthesize more tightly bound species. Many reactions achieve equilibrium during Si burning. In fact, numerical network calculations show that two major groups of nuclides form equilibrium clusters (also called \emph{quasiequilibrium clusters}): one forms around $^{28}$Si (and extends up to $A\approx40$), the other one forms around the iron peak nuclei (starting near $A\approx50$). These two clusters are only weakly linked by other reactions and thus are not in mutual equilibrium for a significant time during Si burning. One major reason is that $^{40}$Ca is a doubly-magic nucleus (with 20 protons and 20 neutrons), so that the capture of a light particle is energetically unfavorable. Hence, the resulting nuclides are quickly photodisintegrated. Nevertheless, given enough time a physical system will seek a state of most favorable energy and, via reactions in the $A\approx40-50$ range, the abundances of the ``Si cluster species'' decline with time in favor of those in the ``iron peak cluster''. Clearly, a plasma composed of iron peak nuclei is energetically much more favorable than one of silicon: binding energies per nucleon near 8.80 MeV are achieved by the iron peak species $^{62}$Ni, $^{58}$Fe and $^{56}$Fe, while the value for $^{28}$Si is only 8.45 MeV. Detailed reaction network calculations show that at the end of Si burning the most abundant product is $^{56}$Fe.

\subsection{Nuclear statistical equilibrium}
As $^{28}$Si disappears in the core at the end of Si burning, the temperature increases until all non-equilibrated reactions come into equilibrium. Now one large cluster stretches from protons, neutrons and $\alpha$-particles all the way to the iron peak and the reaction network attains \emph{nuclear statistical equilibrium} (NSE). The abundance of any nuclide in nuclear statistical equilibrium  can be calculated from a repeated application of the Saha equation. For species $^A_\pi Y_\nu$, with mass number $A$, $\pi$ protons and $\nu$ neutrons, one finds for the number abundance
\begin{equation}
N_Y = N_p^\pi  N_n^\nu  \frac{1}{{\theta ^{A-1}}}\left( {\frac{{M_Y}}{{M_p^\pi  M_n^\nu}}} \right)^{3/2} \frac{{g_Y}}{{2^A}}G_Y^{\mathrm{norm}} e^{B(Y)/kT} 
\label{repeatS}
\end{equation}
where $\theta$ is a constant \cite{Ili07}, $N_p$ and $N_n$ are the number abundances of free protons and neutrons, respectively, $M_i$ is the nuclear mass of species $i$, $g_Y$ is the statistical weight (which depends on the spin of $Y$), $G_Y^{norm}$ is the normalized partition function (which depends on energies and spins of excited levels in $Y$), and $B(Y)$ is the binding energy. Note that in the above equation reaction \emph{rates} are absent, which of course is expected since the reaction network has achieved equilibrium.

Provided that the nuclear physics information on binding energies, spins and excitation energies is available, the abundance of any nuclide in nuclear statistical equilibrium is determined by only three independent parameters: temperature, density and neutron excess. The latter parameter is defined as $\eta \equiv
\sum_i (\nu_i-\pi_i)X_i/M_i$, with $X_i$ the mass fraction. The sum runs over all species $i$ present in the plasma. The neutron excess parameter represents the number of excess neutrons per nucleon and can only change as a result of weak interactions. For example, if only $^4$He, $^{12}$C and $^{16}$O are present in the plasma, then $\eta=0$.

A number of interesting properties can be derived from Eq. (\ref{repeatS}). First, it can be shown that, if we keep the density constant and raise the temperature, an increasing fraction of the composition resides in the light species protons, neutrons and $\alpha$-particles. The importance of this aspect for the core collapse was already mentioned in the introduction. Second, it turns out that the neutron excess parameter influences sensitively the composition during nuclear statistical equilibrium. In fact, NSE favors the abundance of that particular nuclide for which (i) the individual neutron excess is equal to the total neutron excess, \emph{and} (ii) the binding energy is at maximum. For example, when $\eta\approx0$, then $^{56}$Ni (with 28 neutrons and 28 protons; $\eta_i=0$) is the most abundant species. For $\eta\approx0.04$, the most abundant species becomes $^{54}$Fe (with 28 neutrons and 26 protons; $\eta_i=0.037$), and so on. It is evident from this discussion that the neutron excess must be monitored very carefully during all burning stages that precede NSE. Thus \emph{stellar} weak interaction rates must be known reliably (see contribution by G. Martinez-Pinedo in this volume; these rates are also crucial for the core collapse, as already pointed out in the introduction).

\subsection{Explosive silicon, oxygen, neon and carbon burning}
After the collapse of the massive star core, a shock wave is eventually generated and moves outward. It heats and compresses first the innermost layer outside the core, consisting of $^{28}$Si as discussed above,  to high temperature and density. This matter quickly reaches NSE and then expands and cools as the shock wave moves further out. At this point, reactions begin to fall out of equilibrium. In many situations NSE is terminated by an excess of $\alpha$-particles, a process referred to as \emph{$\alpha$-rich freezeout}: the $\alpha$-particles are captured on nuclei and change the composition somewhat from what one would expect from NSE. The main constituent is still $^{56}$Ni (since the $^{28}$Si layer has $\eta=0$; see above), but the abundances of some other iron peak nuclides are modified. The $\alpha$-rich freezeout is also predicted to be the major source of the important radioisotope $^{44}$Ti. It becomes now clear why the light curves of type II supernovae are powered for long time periods by the radioactive decay of $^{56}$Co. The $^{56}$Ni synthesized in the Si layer during the explosion is ejected and decays with a half life of $T_{1/2}=6.1$ d to $^{56}$Co, which itself is radioactive with a half life of $T_{1/2}=77.3$ d. By fitting the light curve of SN 1987A, a $^{56}$Ni mass of about $0.07M_\odot$ has been derived, in reasonable agreement with stellar model predictions.

The outward moving shock wave subsequently reaches the O, Ne and C layers, which were synthesized during hydrostatic burning and are located beyond the silicon layer. Explosive burning in these regions is characterized by temperatures that are somewhat higher than in their hydrostatic counterparts but, nevertheless, the products of nucleosynthesis are very similar. Explosive O burning is predicted to be the main source of the ``$\alpha$ elements'' ($^{28}$Si, $^{32}$S, $^{36}$Ar, $^{40}$Ca etc.). Explosive Ne and C burning, on the other hand, has been suggested \cite{Lim06} as the most probable site for the synthesis of the important radioisotope $^{26}$Al ($T_{1/2}=720000$ y). Gamma-rays from the decay of this nuclide have been observed by several satellite-borne instruments. Since the half life of $^{26}$Al is very short compared to the time scale of Galactic chemical evolution, the observation of ``live'' $^{26}$Al directly demonstrates that nucleosynthesis is currently active in the Galaxy.

Most supernovae ($\approx85$\%) result from the core collapse of massive stars ($>11M_\odot$). A smaller fraction ($\approx15$\%) are of the type Ia variety (see introduction). As already noted, the most popular model  involves the disruption of a C-O white dwarf. Explosive burning of C during the thermonuclear runaway produces mainly $^{56}$Ni via NSE in the hottest zone (since $\eta=0$ for matter consisting of $^{12}$C and $^{16}$O). This provides a natural explanation for the fact that light curves of type Ia supernovae are also powered by the radioactive decay of $^{56}$Co. The outer regions attain lower temperatures and experience explosive Si and O burning, the details of which depend on the explosion mechanism.

\section{Summary}
The origin of the light nuclides ($A\leq40$) is summarized in Tab. \ref{tab:a} (see caption for an explanation of the symbols). As already pointed out, hydrogen ($^{1}$H and $^{2}$H) and helium ($^{3}$He and $^{4}$He) are produced in the Big Bang, while the volatile species $^{6}$Li, $^{9}$Be and $^{10}$B are made chiefly via cosmic ray spallation. Asymptotic giant branch (AGB) stars are the main producer of $^{14}$N and contribue significantly to the cosmic abundances of $^{12}$C, $^{13}$C, $^{22}$Ne, $^{25}$Mg and $^{26}$Mg. Classical novae are predicted to be the main source of $^{15}$N and $^{17}$O, and may also contribute to the abundance of $^{13}$C. All the other nuclides in the $A=16-40$ region are predominantly produced in various hydrostatic and explosive burning stages of massive stars. A few of the light nuclides, most notably $^{7}$Li, $^{11}$B, $^{19}$F, $^{36}$S, $^{37}$Cl and $^{40}$Ar, are of uncertain origin. Beyond $A=40$, the nuclides are either made in massive stars (via $\alpha$-rich freezeout, the weak component of the s-process, the r-process, and the p-process; see contributions by M. El Eid and K. L. Kratz), AGB stars (the main and strong components of the s-process), and in thermonuclear (type Ia) supernovae (about half of the $^{56}$Fe, which is made as radioactive $^{56}$Ni).

We have now reached the end of this brief survey of nuclear astrophysics. The student may note with interest the many areas of nuclear physics that are sampled in stars. It is gratifying to see that, after many decades of research, we are in possession of a remarkably successful theory of stellar evolution and nucleosynthesis. It is equally exciting that there are still many unsolved questions in nuclear astrophysics, some of which may hold important implications for related fields such as cosmology, meteoritics and cosmochronology. Thus the future looks bright for aspiring young minds!

\begin{table}
\begin{tabular}{llllll}
\hline
\tablehead{1}{l}{l}{Nuclide}
  & \tablehead{1}{l}{l}{Origin}
  & \tablehead{1}{l}{l}{Nuclide}
  & \tablehead{1}{l}{l}{Origin}
  & \tablehead{1}{l}{l}{Nuclide}
  & \tablehead{1}{l}{l}{Origin}
  \\
\hline
$^{1}$H	  & BB			  & $^{17}$O	  & CN	                 & $^{30}$S     & C        \\
$^{2}$H	& BB			   & $^{18}$O	& He	                     & $^{31}$P    & C         \\
$^{3}$He	& BB			   & $^{19}$F	& [AGB,...]  & $^{32}$S	  & xO         \\
$^{4}$He	& BB			   & $^{20}$Ne	& C			     & $^{33}$S	  & xO, xNe         \\
$^{6}$Li	& CR		   & $^{21}$Ne	& C			     & $^{34}$S	  & xO         \\
$^{7}$Li	& [BB, AGB, CN] & $^{22}$Ne	& He, AGB	     & $^{36}$S	  & [He(s), xC,...]          \\
$^{9}$Be	& CR		  & $^{23}$Na	& C 			     & $^{35}$Cl	  & xO         \\
$^{10}$B	& CR		  & $^{24}$Mg	& C			     & $^{37}$Cl	  & [xO, He(s),...]        \\
$^{11}$B	& [CR,...]	  & $^{25}$Mg	& C, AGB		     & $^{36}$Ar	  & xO         \\
$^{12}$C	& AGB, He	  & $^{26}$Mg	& C, AGB		     & $^{38}$Ar	  & xO         \\
$^{13}$C	& AGB, CN	  & $^{26}$Al	& xC, xNe		     & $^{40}$Ar	  & [He(s), C,...]         \\
$^{14}$N	& AGB		  & $^{27}$Al	& C			     & $^{39}$K	  & xO         \\
$^{15}$N & CN		        & $^{28}$Si	  & xO 		           & $^{40}$K    & He(s)         \\
$^{16}$O & He		        & $^{29}$Si	  & C   	                  & $^{40}$Ca  & xO    \\
\hline
\end{tabular}
\caption{Origin of the light nuclides. The labels denote: Big Bang (BB); cosmic ray spallation (CR); asymptotic giant branch stars (AGB); classical novae (CN); helium, carbon, neon, oxygen (He, C, Ne, O), where an ``x'' in front of the symbol indicates explosive rather than hydrostatic burning; the weak s-process component is denoted by He(s). Uncertain origin is given in square parenthesis. Information from Refs. \cite{Woo02,Cla03} and elsewhere.}
\label{tab:a}
\end{table}


\begin{theacknowledgments}
This work was supported in part by the U.S. Department of Energy under Contract No. DE-FG02-97ER41041.
\end{theacknowledgments}



\bibliographystyle{aipproc}   

\bibliography{sample}

\begin{thebibliography}{9}

\bibitem{Bet38} H.~A. Bethe, and C.~L. Critchfield, \emph{Phys. Rev.} \textbf{54},
 248 (1938).

\bibitem{Bet39} H.~A. Bethe, \emph{Phys. Rev.} \textbf{55},
 434 (1939).

\bibitem{Wei38} C.~F. von Weizs\"acker, \emph{Phys. Z.} \textbf{39},
 633 (1938).

\bibitem{Bur57} E.~M. Burbidge, G.~R. Burbidge, W.~A. Fowler, and F. Hoyle, \emph{Rev. Mod. Phys.} \textbf{29}, 547 (1957).

\bibitem{Cam57} A.~G.~W. Cameron, \emph{Pub. Astron. Soc. Pac.} \textbf{69}, 201 (1957).

\bibitem{Lod03} K. Lodders, \emph{Astrophys. J.} \textbf{591}, 1220 (2003).

\bibitem{Ili07} C. Iliadis, \emph{Nuclear Physics of Stars}, Wiley-VCH, Weinheim, 2007.

\bibitem{Hir87} K. Hirata, et al., \emph{Phys. Rev. Lett.} \textbf{58}, 1490 (1987).

\bibitem{Mat88} S. Matz, et al., \emph{Nature} \textbf{331}, 416 (1988).

\bibitem{Rie98} A. Riess, et al., \emph{Astron. J.} \textbf{116}, 1009 (1998); S. Perlmutter, et al., \emph{Astrophys. J.} \textbf{517}, 565 (1999).

\bibitem{Ili97} C. Iliadis, \emph{Nucl. Phys.} \textbf{618}, 166 (1997).

\bibitem{Arn96} D. Arnett, \emph{Supernovae and Nucleosynthesis}, Princeton University Press, Princeton, 1996.

\bibitem{CF88} G.~R. Caughlan, and W.~A. Fowler, \emph{At. Data Nucl. Data Tab.} \textbf{40}, 284 (1988).

\bibitem{Ang99} C. Angulo, et al., \emph{Nucl. Phys. A} \textbf{656}, 3 (1999).

\bibitem{Ili01} C. Iliadis, et al., \emph{Astrophys. J. Suppl.} \textbf{134}, 151 (2001).

\bibitem{Ili09} C. Iliadis, et al., \emph{Nucl. Phys. A}, submitted (2009).

\bibitem{Bac69} J.~N. Bahcall, and R.~M. May, \emph{Astrophys. J.} \textbf{155} 511 (1969). 

\bibitem{LUN} H. Costantini, et al., \emph{Rep. Prog. Phys.} \textbf{72}, 086301 (2009). 

\bibitem{LEN} R. Runkle, et al., \emph{Phys. Rev. Lett.} \textbf{94}, 082503 (2005).

\bibitem{Fox05} C. Fox, et al., \emph{Phys. Rev. C} \textbf{71}, 055801 (2005). 

\bibitem{Cha07} A. Chafa, et al., \emph{Phys. Rev. C} \textbf{75}, 035810 (2007).

\bibitem{New09} J. Newton, et al., \emph{Phys. Rev. C}, submitted (2009).

\bibitem{Ili02} C. Iliadis, et al., \emph{Astrophys. J. Suppl.} \textbf{142}, 105 (2002).

\bibitem{Hoy54} F. Hoyle, \emph{Astrophys. J. Suppl.} \textbf{1}, 121 (1954).

\bibitem{Par09} A. Parikh, et al., \emph{Phys. Rev. C} \textbf{79}, 045802 (2009).

\bibitem{Lim06} M. Limongi, and A. Chieffi, \emph{Astrophys. J.} \textbf{647}, 483 (2006).

\bibitem{Woo02} S.~E. Woosley, A. Heger, and T.~A. Weaver, \emph{Rev. Mod. Phys.} \textbf{74}, 1015 (2002).

\bibitem{Cla03} D.~D. Clayton, \emph{Handbook of Isotopes in the Cosmos}, Cambridge University Press, Cambridge, 2003. 

\end{thebibliography}




\end{document}